\documentstyle[epsfig]{mn}
\newif\ifAMStwofonts
\newcommand{\etal}{\mbox{\em et al. }}

\def\kms{\relax \ifmmode {\,\rm km\,s}^{-1}\else \,km\,s$^{-1}$\fi}
\def\ha{\relax \ifmmode {\rm H}\alpha\else H$\alpha$\fi}
\def\hb{\relax \ifmmode {\rm H}\beta\else H$\beta$\fi}
\def\hi{\relax \ifmmode {\rm H\,{\sc i}}\else H\,{\sc i}\fi}
\def\hii{\relax \ifmmode {\rm H\,{\sc ii}}\else H\,{\sc ii}\fi}
\def\h2{\relax \ifmmode {\rm H}_2\else H$_2$\fi}
\def\lha{\relax \ifmmode L_{{\rm H}\alpha}\else $L_{{\rm H}\alpha}$\fi}
\def\shi{\relax \ifmmode \sigma_{{\rm HI}}\else $\sigma_{\rm HI}$\fi}
\def\sh2{\relax \ifmmode \sigma_{{\rm H}_2}\else $\sigma_{{\rm H}_2}$\fi}
\def\degr{\hbox{$^\circ$}}
\def\arcmin{\hbox{$^\prime$}}
\def\arcsec{\hbox{$^{\prime\prime}$}}
\def\deg{\hbox{$^\circ$}}

\def\fdg{\hbox{$.\!\!^\circ$}}
\def\fs{\hbox{$.\!\!^{\rm s}$}}
\def\farcm{\hbox{$.\mkern-4mu^\prime$}}
\def\farcs{\hbox{$.\!\!^{\prime\prime}$}}
\def\degd#1.#2{ #1\fdg#2 }                 % degrees over decimal point
\def\mind#1.#2{ #1\farcm#2 }               % minutes over decimal point
\def\secd#1.#2{ #1\farcs#2 }               % seconds over decimal point
\def\hhh{\ifmmode {\rm ^h}              % hours symbol
         \else {${\rm ^h}$}
         \fi}
\def\sss{\ifmmode {\rm ^s}              % seconds symbol
         \else {${\rm ^s}$}
         \fi}
\def\hms#1h#2m#3s{                      % hms format (for RA)
                  \relax
                  \ifmmode #1^{\rm h}\,#2^{\rm m}\,#3^{\rm s}
                  \else \hbox{$#1^{\rm h}\,#2^{\rm m}\,#3^{\rm s}$}
                  \fi
                 }
\def\dms#1d#2m#3s{                      % dms format (for Dec)
                  \relax
                  #1\degr\,#2\arcmin\,#3\arcsec 
                 }
\def\hmsd#1h#2m#3.#4s{                  % hms format with decimal point (RA)
                      \relax
                      \ifmmode #1^{\rm h}\,#2^{\rm m}\,#3\fs#4
                      \else \hbox{$#1^{\rm h}\,#2^{\rm m}\,#3\fs#4$}
                      \fi
                     }
\def\dmsd#1d#2m#3.#4s{                  % dms format with decimal point (Dec)
                      \relax
                      #1\degr\,#2\arcmin\,#3\farcs#4
                     }
\def\mag{\relax                          % magnitudes symbol
        \ifmmode ^{\rm m}
        \else $^{\rm m}$
        \fi
       }
\def\magd#1.#2{                          % magnitudes over decimal point
              \relax
              \ifmmode #1^{\rm m}
                       \hskip-0.55em.\hskip0.22em#2
              \else \hbox{#1$^{\rm m}
                    \hskip-0.55em.\hskip0.22em$#2}
              \fi
             }

\ifoldfss
  \newcommand{\etal}{\mbox{\em et al. }}

  \ifCUPmtlplainloaded \else
    \NewTextAlphabet{textbfit} {cmbxti10} {}
    \NewTextAlphabet{textbfss} {cmssbx10} {}
    \NewMathAlphabet{mathbfit} {cmbxti10} {} % for math mode
    \NewMathAlphabet{mathbfss} {cmssbx10} {} %  "   "    "
  \fi
  \ifAMStwofonts
    \ifCUPmtlplainloaded \else
      \NewSymbolFont{upmath} {eurm10}
      \NewSymbolFont{AMSa} {msam10}
      \NewMathSymbol{\upi}     {0}{upmath}{19}
      \NewMathSymbol{\umu}     {0}{upmath}{16}
      \NewMathSymbol{\upartial}{0}{upmath}{40}
      \NewMathSymbol{\leqslant}{3}{AMSa}{36}
      \NewMathSymbol{\geqslant}{3}{AMSa}{3E}

    \fi
  \fi
\fi % End of OFSS

\ifnfssone
  \newmathalphabet{\mathit}
  \addtoversion{normal}{\mathit}{cmr}{m}{it}
  \addtoversion{bold}{\mathit}{cmr}{bx}{it}
  \newmathalphabet{\mathbfit} % math mode version of \textbfit{..}
  \addtoversion{normal}{\mathbfit}{cmr}{bx}{it}
  \addtoversion{bold}{\mathbfit}{cmr}{bx}{it}
  \newmathalphabet{\mathbfss} % math mode version of \textbfss{..}
  \addtoversion{normal}{\mathbfss}{cmss}{bx}{n}
  \addtoversion{bold}{\mathbfss}{cmss}{bx}{n}
  \ifAMStwofonts
    \ifCUPmtlplainloaded \else
      \UseAMStwoboldmath
      \makeatletter
      \new@mathgroup\upmath@group
      \define@mathgroup\mv@normal\upmath@group{eur}{m}{n}
      \define@mathgroup\mv@bold\upmath@group{eur}{b}{n}
      \edef\UPM{\hexnumber\upmath@group}
      \new@mathgroup\amsa@group
      \define@mathgroup\mv@normal\amsa@group{msa}{m}{n}
      \define@mathgroup\mv@bold\amsa@group{msa}{m}{n}
      \edef\AMSa{\hexnumber\amsa@group}
      \makeatother
      \mathchardef\upi="0\UPM19
      \mathchardef\umu="0\UPM16
      \mathchardef\upartial="0\UPM40
      \mathchardef\leqslant="3\AMSa36
      \mathchardef\geqslant="3\AMSa3E
    \fi
  \fi
\fi % End of NFSS release 1

\ifnfsstwo
  \DeclareMathAlphabet{\mathbfit}{OT1}{cmr}{bx}{it}
  \SetMathAlphabet\mathbfit{bold}{OT1}{cmr}{bx}{it}
  \DeclareMathAlphabet{\mathbfss}{OT1}{cmss}{bx}{n}
  \SetMathAlphabet\mathbfss{bold}{OT1}{cmss}{bx}{n}
  \ifAMStwofonts
    \ifCUPmtlplainloaded \else
v      \DeclareSymbolFont{UPM}{U}{eur}{m}{n}
      \SetSymbolFont{UPM}{bold}{U}{eur}{b}{n}
      \DeclareSymbolFont{AMSa}{U}{msa}{m}{n}
      \DeclareMathSymbol{\upi}{0}{UPM}{"19}
      \DeclareMathSymbol{\umu}{0}{UPM}{"16}
      \DeclareMathSymbol{\upartial}{0}{UPM}{"40}
      \DeclareMathSymbol{\leqslant}{3}{AMSa}{"36}
      \DeclareMathSymbol{\geqslant}{3}{AMSa}{"3E}
    \fi
  \fi
\fi % End of NFSS release 2

\ifCUPmtlplainloaded \else
  \ifAMStwofonts \else % If no AMS fonts
    \def\upi{\pi}
    \def\umu{\mu}
    \def\upartial{\partial}
  \fi
\fi

\title[NIR Polarized Bipolar Cone in Circinus]{A Near Infrared Polarized 
Bipolar Cone in the CIRCINUS Galaxy}
\author[M. Ruiz et al.]{M. Ruiz$^1$\thanks{email: mili@star.herts.ac.uk}, 
D. M. Alexander$^2$, S. Young$^1$, J. Hough$^1$, S. L. Lumsden$^3$, C.A. 
Heisler$^4$\thanks{We would like to dedicate this work to her memory }
\\ $^1$Department of Physical Sciences, University of Hertfordshire, 
Hatfield, Herts AL10 9AB, UK. 
\\$^2$International School for Advanced Studies, SISSA, Via Beirut 2-4, 
34014 Trieste, Italy. 
\\$^3$Department of Physics and Astronomy, University of Leeds, Leeds 
LS2 9JT, UK.
\\$^4$Mount Stromlo and Siding Spring Observatories, Private Bag,
Weston Creek P.O., Weston, ACT 2611, Australia}

\pagerange{\pageref{firstpage}--\pageref{lastpage}}
\pubyear{2000}
\begin{document}
\setcounter{footnote}{0}
\maketitle
\label{firstpage}
\begin{abstract}
  
We present near--infrared broad--band polarization images of the nuclear
regions of the Circinus galaxy in the J, H and K bands.  For the first time
the south--eastern reflection cone is detected in polarized light, which is 
obscured at
optical wavelengths behind the galactic disk. This biconical structure is
clearly observed in J and H band polarized flux whilst in the K band a more
compact structure is detected. Total flux J--K and H--K colour maps reveal a
complex colour gradient toward the south--east direction (where the
Circinus galactic disk is nearer to us). We find enhanced extinction in an
arc shaped structure, at about 200pc from the nucleus, probably part of the
star-formation ring.

We model the polarized flux images with the scattering and torus model of
Young \etal, with the same basic input parameters as used by Alexander \etal 
in the spectropolarimetry modelling of Circinus.
The best fit to the polarized flux is achieved with a
torus radius of $\sim$16pc, and a visual extinction A$_V$, through the torus,
to the near--infrared emission regions of
$>$66 mags.

\end{abstract}

\begin{keywords}
galaxies: active -- 
galaxies: individual (Circinus) -- 
galaxies: nuclei -- 
galaxies: Starburst -- 
infrared: galaxies
\end{keywords}

\section{Introduction}

\begin{figure*}
%  \begin{center}
%    \leavevmode
\centerline{\epsfig{file=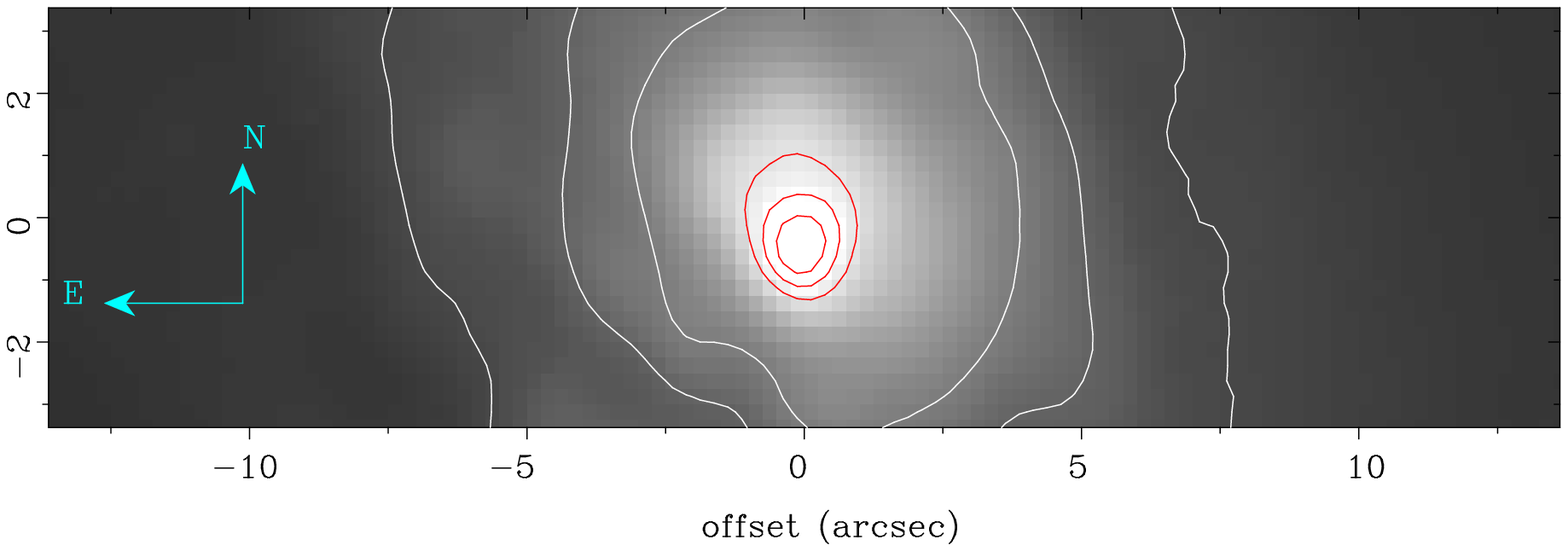,width=14cm,height=4cm,angle=0}}
\vspace{0.4cm}
\centerline{\epsfig{file=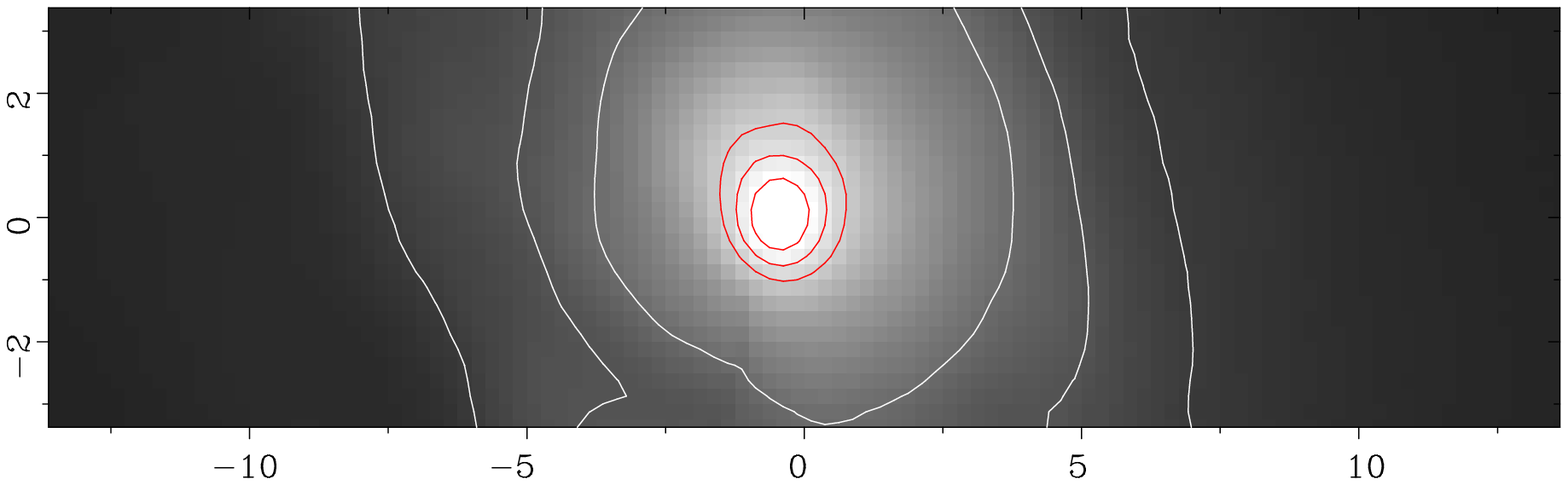,width=14cm,height=4cm,angle=0}}
\vspace{0.4cm}
\centerline{\epsfig{file=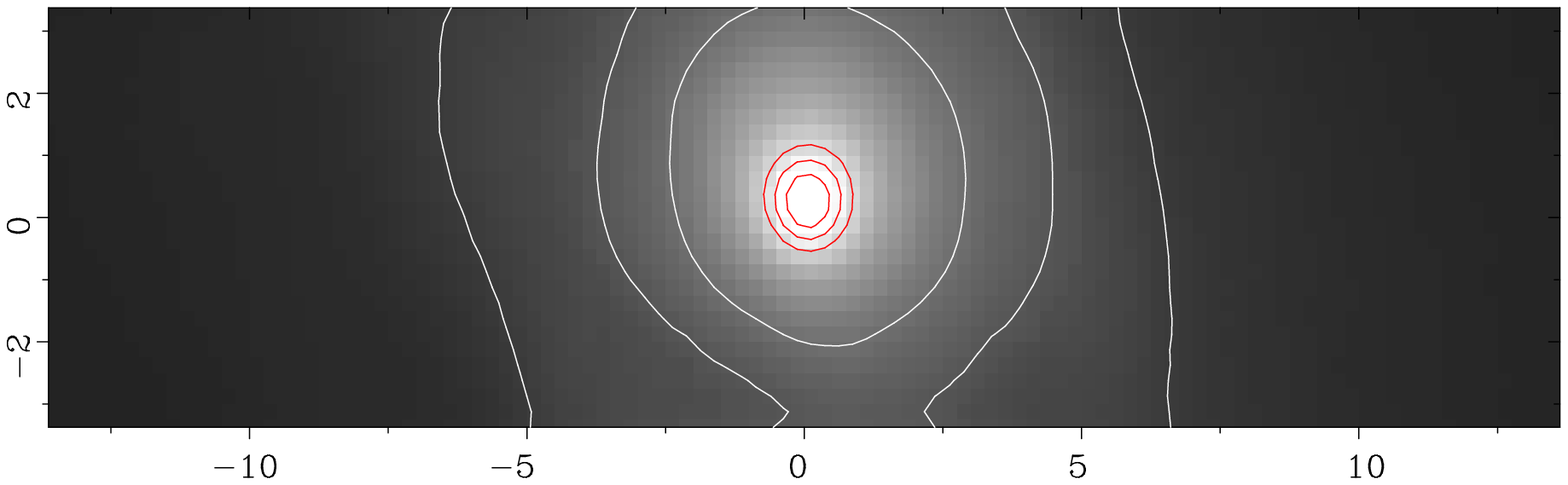,width=14cm,height=4cm,angle=0}}
\caption{Near-IR images of the Circinus galaxy, J band (top), 
H band (middle) and K band (bottom). Contours are arbitrarily scaled.}
  \label{}
\end{figure*}

 The investigation of nearby Seyfert galaxies with 
polarimetric observations provides
unique information on the nuclear  
structure of these objects. This information is of  
particular importance when studying unification models of Seyfert galaxies.

\begin{figure*}
\centerline{\epsfig{file=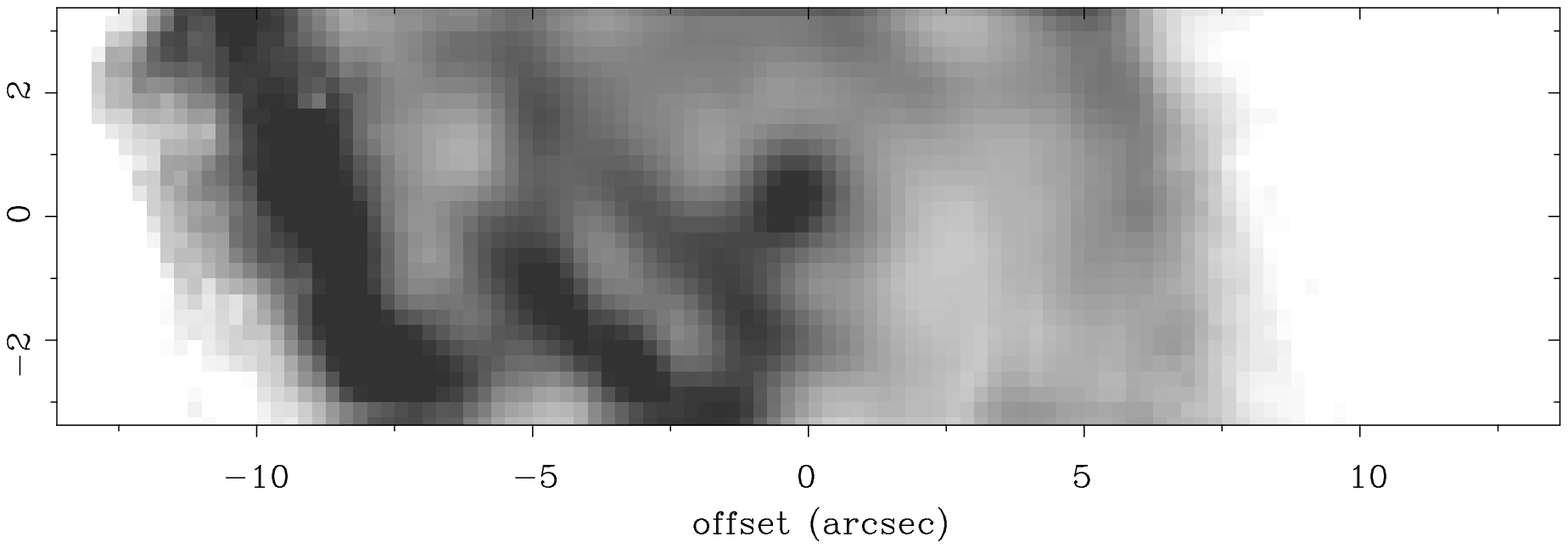,width=14cm,height=4cm,angle=0}}
\vspace{0.4cm}
\centerline{\epsfig{file=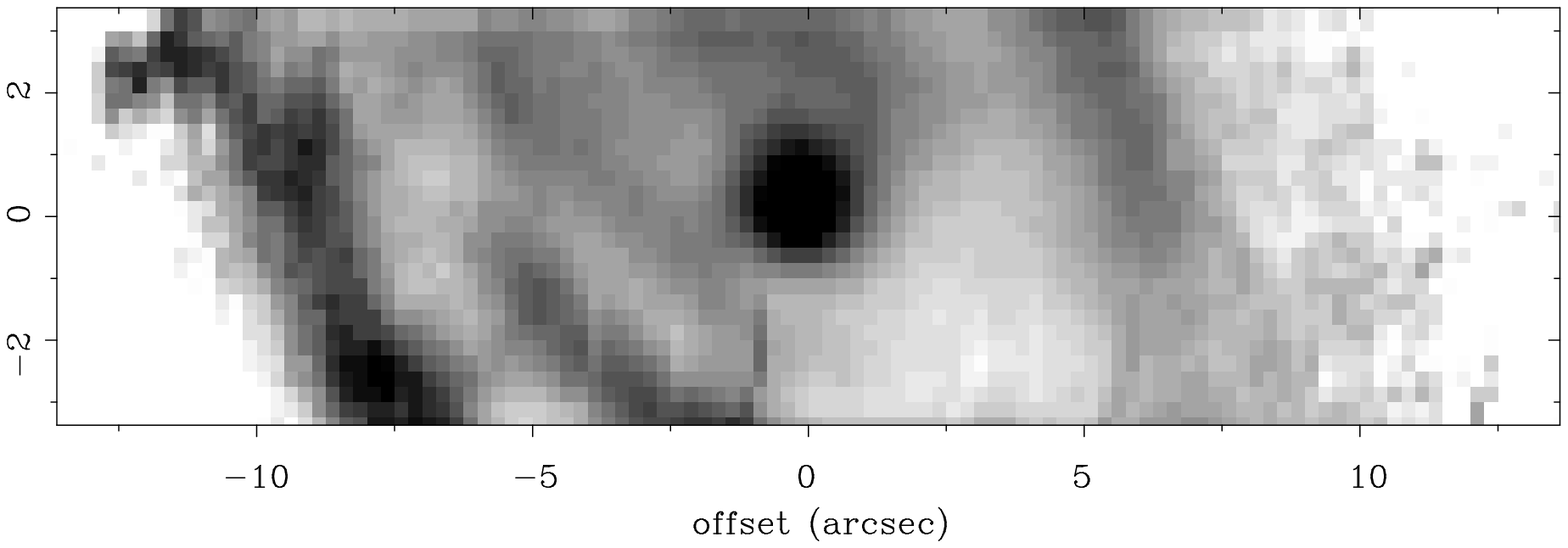,width=14cm,height=4cm,angle=0}}
\vspace{0.4cm}
\caption{Colour maps of the nuclear regions of the Circinus galaxy. J--K  
(top) and H--K (bottom). The dark areas correspond to regions of enhanced 
extinction.}
\label{}
\end{figure*}

The standard unified model for Seyfert galaxies proposes that all types of
Seyfert galaxy are fundamentally the same, however, the presence of a dusty
molecular ``torus'' obscures the broad line emission in many systems. In
this picture the classification of Seyfert 1 or 2 depends on the
inclination angle of the torus to the line of sight (Antonucci, 1993). The
most convincing evidence for this unified model comes from optical
spectropolarimetry. Using this technique, the scattered radiation from the
broad line region (BLR) of many Seyfert~2 galaxies is revealed in the form
of broad lines in the polarized flux (e.g.\ Antonucci and Miller, 1985,
Young \etal, 1996a, Heisler, Lumsden and Bailey, 1997).

Near--IR imaging polarimetry provides valuable information on the nature of
the polarizing source (e.g.\ Lumsden \etal, 1999, Tadhunter \etal, 1999,
Young \etal, 1996b, Packham \etal, 1996, 1997, 1998, 1999). For example, in
NGC1068, bipolar scattering cones are clearly detected in polarized flux
(Young \etal 1996b, Packham \etal 1997) and the torus itself has been
viewed in silhouette in the H band (Young \etal 1996b). Interestingly, the
structure of the scattering cones often coincide with ground-based 
narrow band imaging (e.g.
Wilson and Tsvetanov, 1994) and high resolution HST imaging (e.g. Capetti
\etal 1997, Falcke, Wilson and Simpson, 1998). Since the unified model
infers the presence of a dusty torus obscuring the Seyfert 1 core, we
expect that longer wavelength polarimetry will be able to probe deeper into
regions within the plane of the torus and to see scattering from the nucleus
which might otherwise be shielded from view at optical wavelengths.
  
Circinus is a nearby (4Mpc) highly inclined (65$\deg$, Freeman \etal 1977)  
Sb-Sd galaxy. At this distance, the spatial scale is 20pc/arcsec. 
It is seen through a low interstellar extinction window near
the Galactic plane (A$_V$ = 1.5 mag; Freeman \etal 1977). The nuclear
optical line ratios are typical of a Seyfert 2 galaxy. This classification
as a type 2 is also supported by the detection of intense coronal lines
(Oliva \etal 1994), the intense X-ray Fe 6.4 keV line (Matt \etal 1996),
rapid variation of powerful H$_2$O maser emission and a prominent
ionization cone in [O\,{\sc iii}]5007 with filamentary supersonic outflows
(Marconi \etal 1994). Recent optical spectropolarimetry (Oliva \etal, 1998
and Alexander \etal, 1999a) has shown a scattered polarized broad H$\alpha$
line and therefore a hidden Seyfert 1 nucleus. Other characteristics
include enhanced star forming activity in the form of a star-forming ring
of 200pc in size (Marconi \etal 1994), a Compton thick nucleus at X-ray
energies (Matt \etal 1996) and an unresolved nuclear source ($<$1.5pc) at
2$\mu$m (Maiolino \etal 1998).

Since it is already known from optical spectropolarimetry that the Circinus
galaxy harbours a hidden type 1 nucleus and the optical ionization cone
suggests the presence of an obscuring/collimating torus, this galaxy is an
ideal candidate to test the theoretical models of unified schemes. The
dusty nature of the Circinus galaxy favours near-IR polarimetry over
optical polarimetry due to the lower optical depth at near-IR wavelengths.

In this paper we present near--IR imaging polarimetry revealing, for the
first time, a biconical emission region. We investigate the nature of this
emission in the context of the standard unified model of Seyfert galaxies.

\begin{figure*}
%  \begin{center}
%    \leavevmode
%  {file=j.ps,width=3cm,height=14cm,angle=-90}
\centerline{\epsfig{file=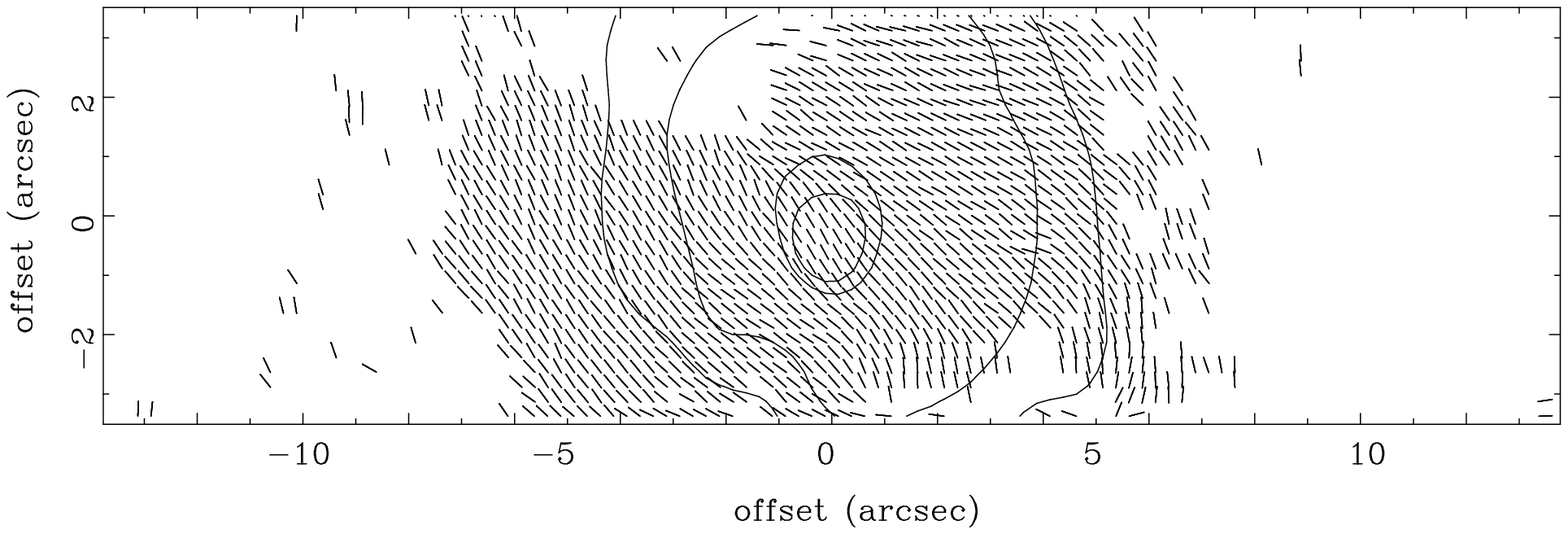,width=14cm,height=5cm,angle=0}}
\vspace{0.4cm}
\centerline{\epsfig{file=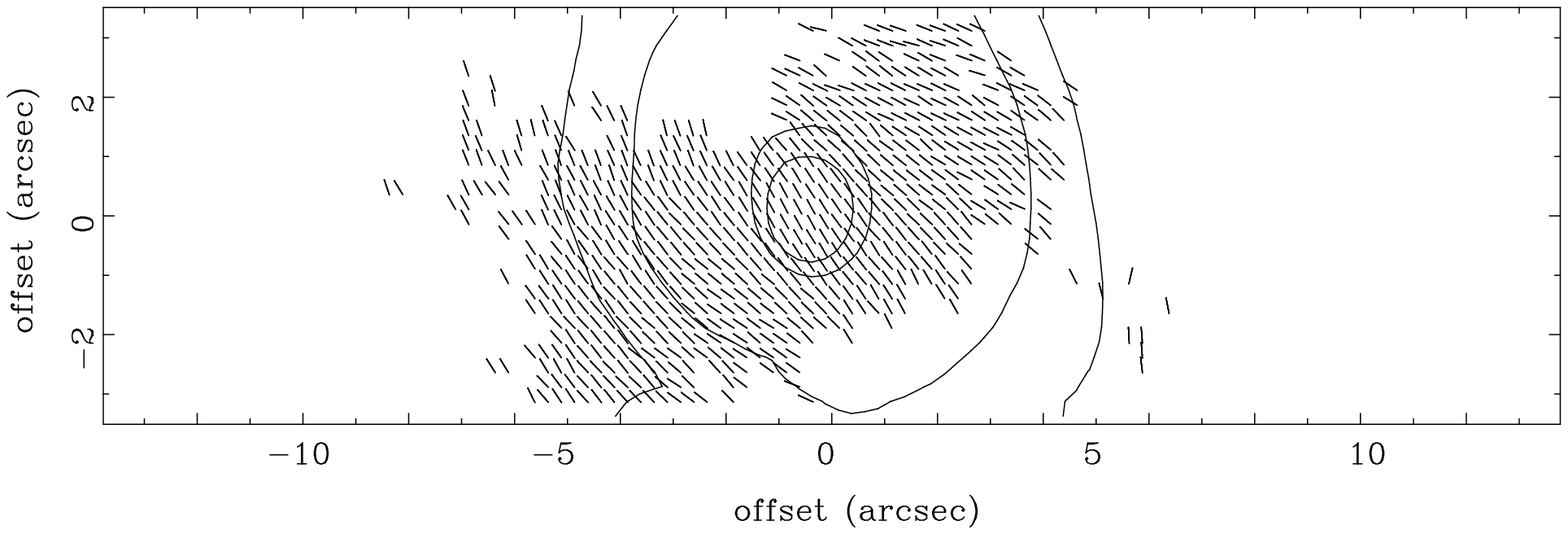,width=14cm,height=5cm,angle=0}}
\vspace{0.4cm}
\centerline{\epsfig{file=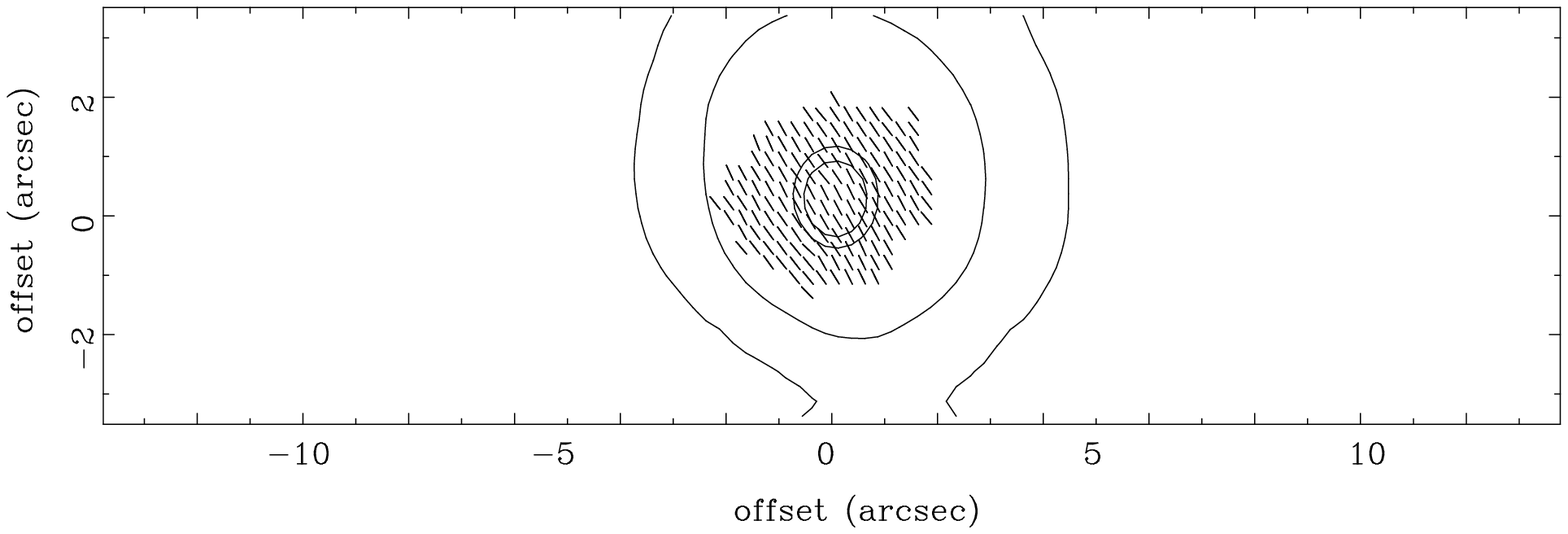,width=14cm,height=5cm,angle=0}}
  \caption{Near-IR polarization maps of Circinus. The J band data is shown
at the top, 
H band in the middle and K band at the bottom. The contours are of the
total flux and are arbitrarily scaled. A 1 arcsec polarization vector 
corresponds 
to a 10 percent polarization.}

\label{}
 \end{figure*}

\begin{table}
\begin{minipage}[t]{5.5in}
  \caption{Nuclear polarization}
  \label{tab:table}
    \leavevmode   
    \footnotesize
    \begin{tabular}[h]{cccc}
%      \hline \\[-5pt]
Waveband & aperture & $\%$ &angle\\
         &  (arcsec)& & ($\deg$)\\
%      \hline \hline \\[-5pt]
        &	&                    &             \\
        &    1   &  2.04$\pm$0.38    &  34.1$\pm$3.8\\
J        &    2   &  1.71$\pm$0.16   &  38.3 $\pm$2.0    \\
         &    4   &  1.40$\pm$0.08   &  46.3$\pm$1.1 \\  
        &	&                    &             \\
        &    1      &  2.01$\pm$0.15 &  33.4$\pm$2.4   \\
H        &    2      &  1.75$\pm$0.13 &  35.5$\pm$1.9   \\
	&     4      &   1.34$\pm$0.07    & 41.1$\pm$1.1 \\  
        &	&                    &             \\
        &    1       &   3.25$\pm$0.27  &   32.1$\pm$3.1 \\
K       &    2      &  2.78$\pm$0.25 &  33.7 $\pm$2.9  \\
	&     4    &   1.99$\pm$0.15 &  34.7 $\pm$1.8       \\
%      \hline
      \end{tabular}
\end{minipage}
\end{table}

\section{Observations}

The data presented here were obtained on the nights of 22 May 1995 (H and K
band data) and 11 August 1995 (J band) on the AAT with the common user camera
IRIS, which uses a 128$^2$ HgCdTe array.  
We used the cs/36 secondary, which results in a pixel scale of 
approximately 0.25 arcsec/pixel.
The May observations were made under
non-photometric conditions. Seeing as estimated in the infrared was $1-1.2$
arcsec for the May data, and $\sim0.9$ arcsec for the August data.  The IRIS
photometric standard star SA94-251 was observed at J for the August data.  No
photometric standards are available for the May data given the conditions.
Instead, we used a spectrum of the nucleus of Circinus, taken on 21 Feb 1997 
with the
echelle grisms inside IRIS, to achieve an overall flux calibration as described
below.  The stability and instrumental polarization of the instrument were
checked using polarized and unpolarized standard stars.  The measured
instrumental polarization is less than 0.1$\%$, and the instrument is almost
completely stable between the two dates when data were taken.

The polarimeter inside IRIS uses a Wollaston beam splitting prism inside the
dewar to separate the $o-$ and $e-$rays.  A mask in the focal plane prevents
the separate images from overlapping.  The mask when used with the cs/36
secondary has dimensions on the sky of approximately 
30arcsec$\times$8arcsec.  A $\lambda/2$ waveplate is positioned in front of 
the dewar.  Each
polarimetry dataset is then comprised of exposures at four separate waveplate
positions ($0^\circ, 45^\circ, 22.5^\circ$ and $67.5^\circ$).

The polarimetry data were reduced as follows. All frames were flatfielded using
dome flats. Offset sky frames were taken with varying positions from the
nucleus.  These frames were median filtered to remove background sources,
leaving four separate sky frames for each waveband, corresponding to the four
waveplate positions.  The four sky frames were scaled to the median level of
the sky within each group of four individual object frames, and the result
subtracted from the object frames. The resultant images were then registered
and combined into separate mosaics for each wave plate position. These final
mosaics were combined to form the Q and U Stokes images (using the TSP package
and the ratio method -- Bailey 1997) and hence polarization maps.
Total on-source exposure times for each waveplate position are 
960 seconds at J, and 300 seconds at both H and K.

As noted above, the May data were obtained under poor conditions.  Therefore, 
we did not
attempt to perform an absolute flux calibration for this data therefore.
Instead, we derived relative photometry from the flux-calibrated 1--2.4$\mu$m
spectrum.  This calibration is performed by scaling the J total flux image
counts in a circular aperture of equivalent area to that of the spectrum
aperture (2arcsec$\times$2arcsec).  These counts correspond to the flux as
measured in the spectrum at J. The count ratios H/J and K/J as measured from
the total flux images are then scaled to the corresponding ratios measured from
the spectrum.
The main error in this process is in the limited accuracy with which
the overall spectral shape is defined.
We therefore adopt a
conservative error estimate of 30$\%$ for the overall flux calibration
taken from this data.  

\section{Results}

In this section we present the general results from our data and provide an 
overview 
of the features observed in the central regions of the Circinus galaxy. In 
Fig. 1
we show the J, H and K total flux images of Circinus. 

\subsection{Colour maps}

The J--K and H--K maps are shown in Fig. 2. These maps were generated  by 
registering the peaks of the total flux images, whose relative offsets did not
exceed 2 pixels in RA or DEC. Darker areas correspond to redder colours.

We observe a complex colour structure clearly indicating a non--uniform 
distribution of dust in the Circinus galactic disk. 

In particular, we observe a colour gradient towards the south--east region 
which
is due to an increase in extinction. As discussed by Quillen \etal (1995),
the closest side of a galaxy disc undergoes the most efficient dust screening,
theu, the south-east region of the Circinus galactic disc is closest to us 
(see also section 4.2). 

As previously observed by Maiolino \etal (1998), we detect a dust lane close 
to the nucleus in a
bar--like structure, running from the NE to SW which is also connected 
to the nucleus,
best seen in the J--K map.  There is another dust lane
in a bar--like structure, running parallel to the latter. 
At about 180pc from the nucleus (9 arcsec), to the East, there is an arc of 
heavy extinction, which is likely to 
represent part of the star formation ring previously observed 
in an H$\alpha$ image 
(Marconi \etal 1994).

\begin{figure*}
%  \begin{center}
%    \leavevmode
%  {file=j.ps,width=3cm,height=14cm,angle=-90}
\centerline{\epsfig{file=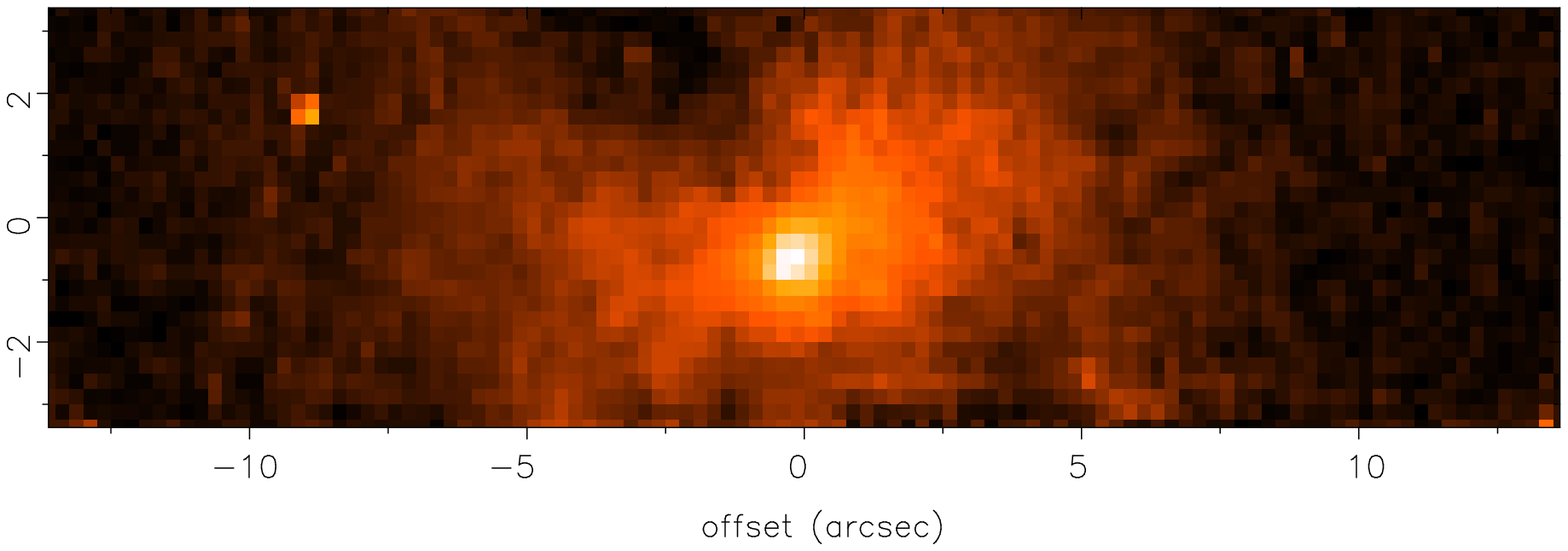,width=14cm,height=4cm,angle=0}}
\vspace{0.4cm}
\centerline{\epsfig{file=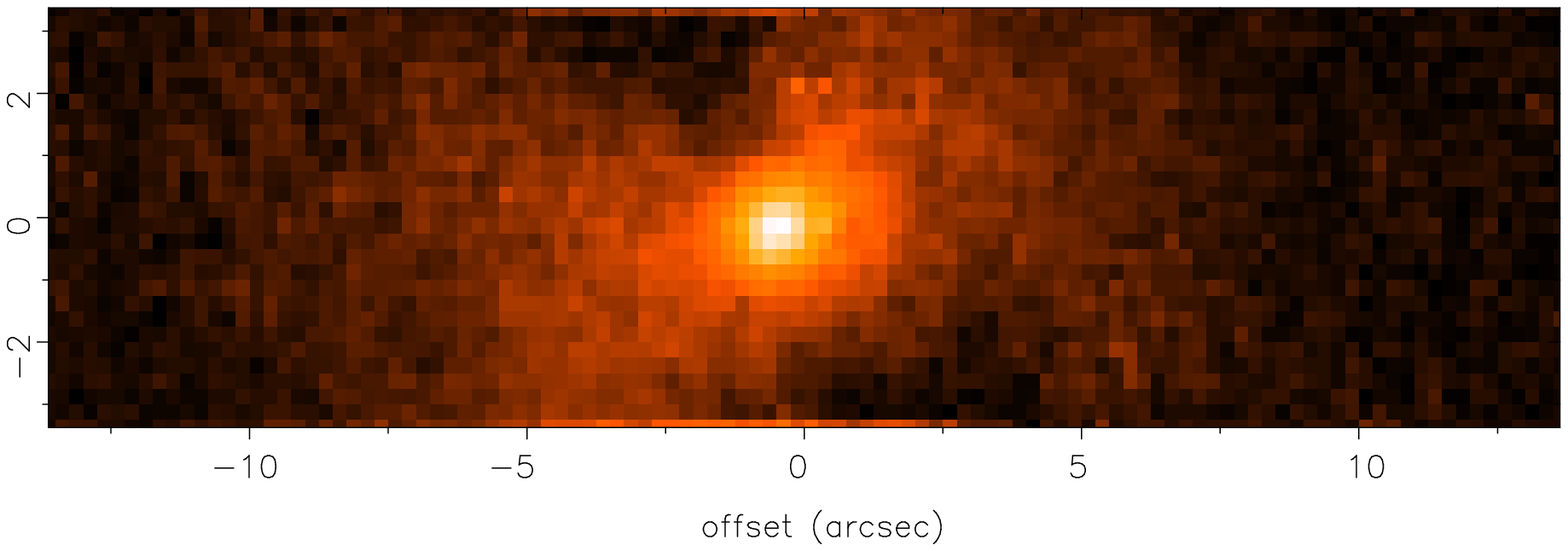,width=14cm,height=4cm,angle=0}}
\vspace{0.4cm}
\centerline{\epsfig{file=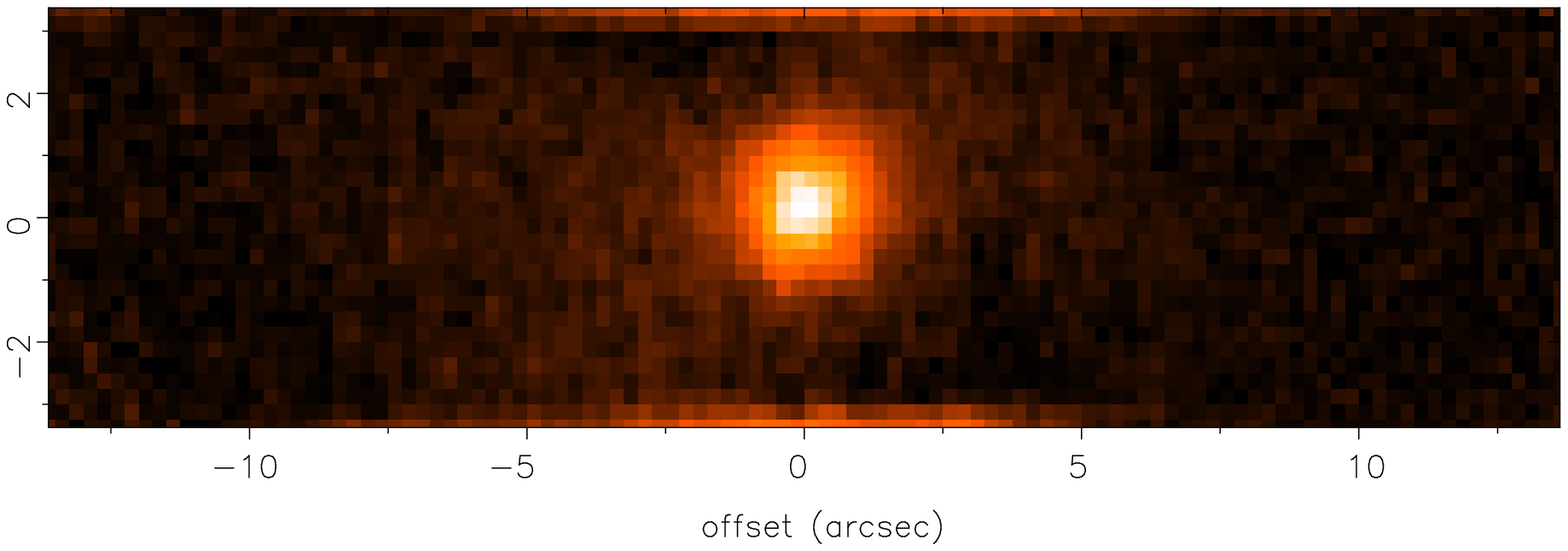,width=14cm,height=4cm,angle=0}}
  \caption{Near-IR polarized flux images of Circinus, J band (top), 
H band (middle) and K band (bottom), showing the change in polarized 
structure from 
predominantly scattering to predominantly dichroism.}
  \label{}
\end{figure*}

\subsection{Polarized maps}

The polarization vector maps are shown 
in Fig. 3, superimposed are the continuum contours. Only those pixels with
a level of at least 3$\sigma$ above the background of polarized intensity 
are displayed. 
The J and H polarization vector maps are similar to each other but there is 
no obvious large--scale symmetry. The highest polarization for J and H is 
within
the 2 arcsec of the nucleus; at J it has a maximum value of 2.04$\pm 0.38\%$
at a position angle of 34\deg$\pm$ 3.8\deg and at H it has  a 
maximum of 2.01$\pm 0.15\%$ with a position angle of 
33.4\deg $\pm 2.4$\deg. The E vectors at K are essentially parallel 
over the central 5 arcsec with a highest 
polarization of $\sim$ 3.25$\pm 0.27\%$  at the  nucleus a
32\deg$\pm 3$\deg and $\sim$ 1.6$\%$ elsewhere. Table 1 presents 
polarization and position
angle values as measured in various aperture sizes. 
We notice that the polarization position 
angle for a nuclear aperture of 1 arcsec is roughly perdendicular 
to the axis of the infrared 
scattering cones seen in the J band and the one--sided optical ionization cone 
(Marconi \etal 1994) as well as to the the radio 
continuum  emission (Elmouttie \etal 1998). 
At larger apertures, the polarization position angle changes, 
due to different contributions
to the total polarization such as galactic polarization. The largest 
change in polarization position angle occurs in the J band. This is 
because the larger 
aperture will have contributions from regions dominated by scattering 
polarization 
i.e., from the scattering cones.
At K, the polarization is dominanted by dichroism (see Section 4) 
at small and large apertures and the small change in the polarization 
position angle is likely to be due to galactic polarization.

\subsection{Polarized flux images}

The near--infrared polarized--flux images are shown in Fig. 4.
The J band polarized image shows a double sided cone--like structure,
with  axis approximately along the NW--SE direction. 
Previous optical imaging of this 
galaxy (Marconi \etal 1994) showed a one-sided ionization cone in the light
of the  [O\,{\sc iii}]5007 line
in the north--west direction. More recently, (Maiolino \etal 1999)  have shown
an extended [Si\,{\sc vi}]1.95$\mu$m  emission in the south--eastern 
region of the nucleus, 
providing evidence for the existence of a counter--cone, whose existence was 
previously inferred from radio maps (Elmouttie \etal 1995, 1998). 
Additionally, the presence of
polarized emission to the south--east of the nucleus 
(the ``counter--cone'') is also indicated 
by the detection of polarized H$\alpha$ emission at about 8arcsec 
to the SE (Oliva \etal 1998).
 
The north--west polarization cone is presumably
produced by scattering in a region spatially coincident with the 
north--west ionization cone. The lack of an ionization cone to the 
south--east is due to the heavy extinction at optical wavelengths 
caused by the galaxy disk ($\imath \simeq 65\deg$), 
estimated at 5 mags in the visible (Oliva \etal 1995; Maiolino \etal 1998). 
In addition, there is evidence for the 
presence of  dust lanes in the south--east direction (Marconi \etal 1994; 
Maiolino \etal 1998) and as shown in Fig 2.

As the wavelength of the observations increases, at H and K bands, 
the bipolar pattern tends to disappear as the amount of scattered 
light reduces, as would
occur for scattering from small dust grains, and with more nuclear 
light being seen directly.
Fig. 5 shows cuts through the nucleus along the east--west direction of
the polarized flux images at J and K, showing the nuclear concentration 
of the K emission to the central 4 arcsec compared to the  
more extended J emission.

\begin{figure*}
\centerline{\epsfig{file=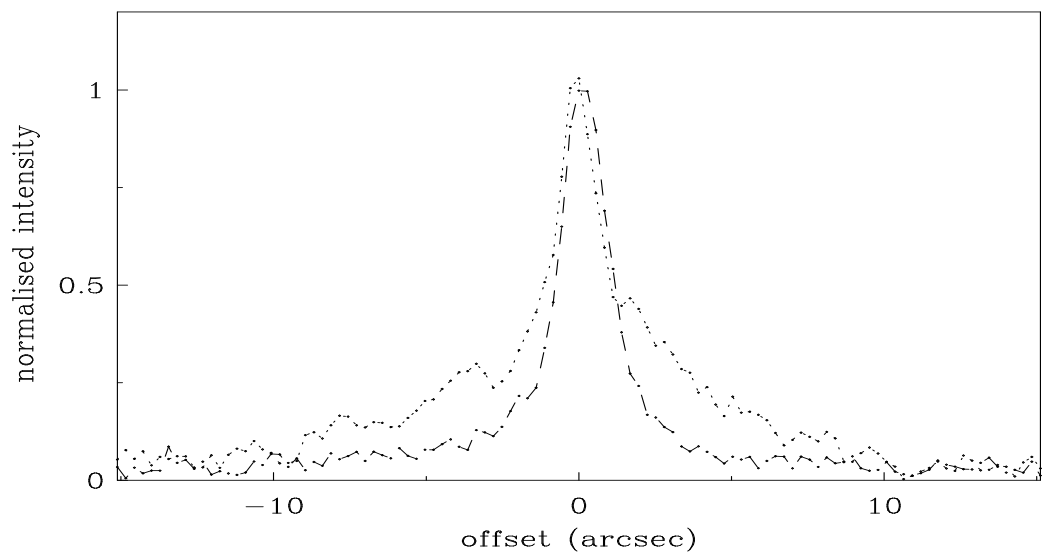,width=14cm,height=8cm,angle=0}}
\vspace{0.4cm}
\caption{Polarized intensity cuts through the nucleus along the 
east--west direction for 
J and K bands. The dashed line represents K and the dotted line 
represents J. The fluxes 
have been normalized to unity.
}
  \label{}
\end{figure*}

\section{The scattering model}

To model
the observed polarization
characteristics of the Circinus galaxy, we have used the standard 
Seyfert model of Young \etal (1995), hereafter referred to as the Y95 model.

\subsection{Description of  the model}

The Y95 model takes the standard unification of AGN approach, assuming a
bare Seyfert~1 nucleus as the source function for the central source, in
this case NGC5548, surrounded by an optically thick torus.  The nuclear
radiation is collimated by the torus and scattered in a biconical
cloud of electrons and/or dust grains.  
The model also considers the direct view to the emission region, that can
be polarized via dichroic absorption by aligned grains within the torus.
The model was originally developed to reproduce
spectropolarimetric data, and was used to successfully model the
polarization of NGC1068 (Young \etal 1995), other narrow line active
galaxies (e.g Young \etal 1996a), and the Circinus galaxy 
(Alexander \etal 1999a).  The model was modified to produce images 
by integrating the
scattered intensity over the size of a pixel at the distance of the galaxy
in question, as illustrated for NGC1068 (Packham \etal 1997).

The most important parameters for the model are the inclination of the
cone axis to the line of sight (also the polar axis of the torus in this
simple model), the cone half opening-angle, the extinction through the
torus to the emission regions and the optical depth to scattering in the
cones.  The latter is defined in terms of inner and outer scattering
radii, a number density of scatterers at the inner radius and the radial
dependence of the number density of scatterers. In the case of spatially
resolved imaging the inner scattering radius and the radial dependence for
the number density also determine the radial scattered light profile with
distance from the nucleus.

\subsection{Applying the model to the Circinus galaxy}

To reduce the number of free variables in the model we can fix some of these
parameters.  As in the spectropolarimetry modelling of Circinus (Alexander
\etal 1999a), we set the cone half opening-angle to 45\deg, which is
similar to the observed  [O\,{\sc iii}]5007 emission line cone (Marconi \etal
1994).  In order to reproduce the intrinsic scattered degree of
polarization at 26$\%$ (Oliva \etal 1995; Alexander \etal 1999a) the
inclination of the scattering axis to the line of sight is 50\deg.
Except for assuming a large outer scattering radius of 1$\times 10^{20}$m, no
other assumptions were made prior to the modelling.

The model output images were smoothed with a Gaussian filter with a FWHM
chosen to match the seeing of the observations, 1 arcsec for J and H and
1.5 arcsec for K, and then scaled by a factor to match the peak of the
polarized image at J.  As previously mentioned in section 2, 
the images are not absolutely flux
calibrated but are in suitable units to represent relative fluxes, thus
the same scaling factor is used at all wavelengths.  Comparisons of the
model output with the observations are then made by producing
cross--section cuts through the images in the SE--NW direction, in polarized
flux.  Modelling the polarized flux images, rather than the total flux, to
the first order removes the stellar population and the uncertainty of the
stellar fraction.

Cross--sections through the observation images, however, show that the
polarized nuclear flux has an underlying base, presumably resulting from
dichroically polarized stellar emission.  This base was fitted separately
by a simple addition to the model output.  At J this base is consistent
with a constant value over the model image size, whilst at H and K the
excess was taken as a linearly dependent ramp across the cut.  With this
pedestal taken into account it is possible to determine the extinction to
the scattering regions away from the nucleus.  For the south--east
 scattering cone
this is consistent with a visual extinction of A$_V$ =~5 mags, while for 
the north--west cone away from the nucleus, 
the extinction is A$_V$ = 1.5 mags.  The
latter value of extinction is the same as that determined for the Galaxy
(Freeman \etal 1977), implying that the south--east scattering cone is actually
viewed through an extinction of A$_V$ = 3.5 mags arising from the Circinus host
galaxy. This value is in agreement with the variation of extinction along
the cone--axis measured by Oliva \etal (1999), as derived from optical 
spectral lines.  

Two possible orientations for the model were investigated, the first has
the less obscured north--west scattering cone pointing towards 
the observer and the
south--east scattering region as the counter--cone, and the second model is the
reverse with the forward pointing cone being to the south--east.  However, 
no fit
to the observations could be found using the first model and therefore,
this case will not be discussed further.  Henceforth, the south--east
scattering cone is considered to be forward pointing.

To match the radial distribution of the scattered flux from the nucleus,
the best fit was obtained with an inner scattering radius of 0.5pc and a
radial dependence for the number density of scatterers as $r^{-1}$.  This
appears to be tightly constrained, with only a 10 percent alteration in
the inner radius being inconsistent with the observations.  
Unlike NGC1068 (Packham \etal 1997),
the dichroically polarized direct view to the near--infrared emission region
is only readily apparent in the K--band image, which suggests that the
extinction through the postulated torus is higher than the A$_V$ = 37 mags
for NGC1068.  However, because we only have one measurement of the direct
view, it is not possible to determine absolutely the extinction through
the torus for the Circinus galaxy.  It is possible to derive a lower limit
for this extinction in conjunction with an upper limit for the number
density of scatterers.  A visual extinction of A$_V$ = 66 mags was
found to be the lowest compatible with the observations, and the upper
limit for the required number density of the scattering electrons at the
inner scattering radius were 3 $\times 10^9$ m$^{-3}$ for the forward 
cone and 4.2 $\times 10^9$ m$^{-3}$ for the counter--cone.  
It should be noted that scattering using
Rayleigh--type dust grains did not provide a good fit to the observations.

To fully match the cross--sectional cuts through the observations it was
necessary to invoke extinction of the scattered flux from the
counter--cone by the torus, which was assumed to be an opaque disk.  Also,
it was found that the best match to the observations was achieved if the
galactic extinction of the forward cone extended partially across the
counter--cone and dropped off with distance away from the nucleus. The
radius of the torus was found to be 16pc, and was well constrained, a
smaller torus being inconsistent at J and a larger torus was inconsistent
at H.  The galactic extinction tail--off was modelled as a simple step,
with the full extinction of A$_V$ = 3.5 mags stepping down to A$_V$ = 2.3 mags
to a distance of 25 pc beyond the nucleus in addition to the Galactic
extinction of 1.5 mags.

The model parameters are listed in Table 2.  Comparison of the
cross--sections through the model produce images and the observations in
polarized flux are illustrated in Fig. 6 for the J, H and K bands.

\begin{figure*}
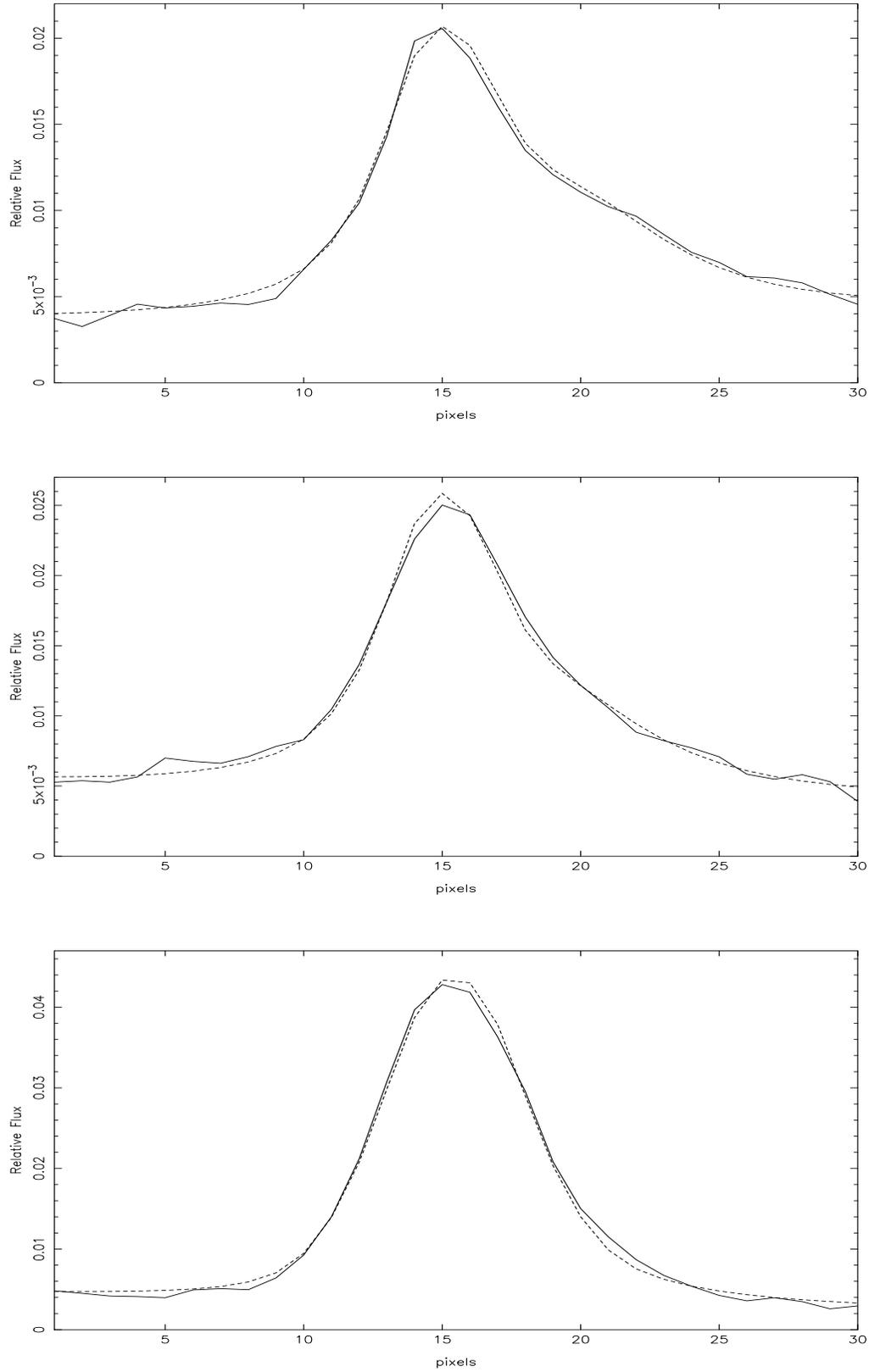


\centerline{\epsfig{file=figure6a.ps,width=7cm,height=14cm,angle=-90}}
\vspace{0.4cm}
\centerline{\epsfig{file=figure6b.ps,width=7cm,height=14cm,angle=-90}}
\vspace{0.4cm}
\centerline{\epsfig{file=figure6c.ps,width=7cm,height=14cm,angle=-90}}
\vspace{0.4cm}
\caption{Cross-sections of polarized flux images along the NW--SE direction.
J band at the top, H band in the middle and K band at he bottom. 
The solid line represents the data and the dashed line is the model fit.
}
\label{}
\end{figure*}

\begin{table}
\begin{minipage}[t]{5.5in}
  \caption{Model parameters.}
  \label{tab:table}
    \leavevmode   
    \footnotesize
    \begin{tabular}[h]{lc}
%\hline \\[-5pt]
\hspace{1.2in}
\vspace{0.1in}
	Input parameters\\
      system inclination($\deg$)          & 50 \\
      SE or NW cone opening half-angle($\deg$)   &  45\\
      SE cone position angle($\deg$)       &125\\
      NW cone position angle($\deg$)       &305\\
                       &          \\
%\hline \\[-5pt]
\hspace{1.0in}
\vspace{0.1in}
	Best fitting parameters\\
      inner scattering radius (pc)  & 0.5\\
      n$_{iscr}$(m$^{-3}$)\footnote{number density at inner radius}     &$< 3 \times 10^9$\\
      n$_{cc}$(m$^{-3}$)\footnote{number density at counter cone}     &$< 4.2 \times 10^{9}$\\
      $\alpha$, power law decrement\footnote{for number density $\sim r^{\alpha}$}           &-1\\
      torus radius (pc)                & 16\\
      A$_V$ through torus           & $>$ 66\\
      A$_V$ through galactic disk             & 3.5\\
      A$_V$ through Galactic disk             & 1.5\\
%      \hline
      \end{tabular}
\end{minipage}
\end{table}

\subsection{Discussion}
 
Alexander \etal (1999a) modelled the optical and K--band spectropolarimetry 
data of the
Circinus galaxy with scattering, and a dichroic component through the dusty
torus corresponding to a visual extinction of A$_V$ = 35 mags.  In the present 
study, we have
the advantage of being able to constrain the scattered flux with both the
spatially resolved information of the images and the extra wavelength
coverage with the J and H band images.  From this we determine that the
ratio of scattered intensity to that from the direct view is higher than
that found by Alexander \etal (1999a), which explains the difference in
the modelled extinction.

We achieved a good fit to our polarized images with a torus radius of
approximately 16~pc.  This is substantially smaller than the size of the
torus determined for NGC1068 of $\sim$ 180 pc (Efstathiou, Hough \& Young 1995;
Young \etal 1996b; Packham \etal 1997), but larger than that estimated for 
Cen A, $\sim$2pc (Alexander \etal 1999b). 

Modelling the polarized flux images of Circinus only allowed us to place an
upper limit for the electron scattering number density, together with a
lower limit for the visual extinction through the torus to the near-IR
emission region.  However, the lower the scatterer's number density, the
higher the boost factor (the ratio of the actual luminosity of the source
to the observed polarized luminosity).  Scattering number densities
significantly less than the upper limits, greater than a factor of 10--20,
result in a boost factor that, using the scattered broad \ha~flux of
4.3$\times 10^{-15}$ erg s$^{-1}$ cm$^{-2}$ (Alexander \etal 1999a), 
implies a broad \ha~luminosity in the top 40 percent of all Seyfert galaxies,
 while
its infrared luminosity is in the lower 20 percent.  This argues that the
number density of scatterers must be within a factor of 10 of the upper
limit.

\section{Conclusions}

We have presented near--infrared polarimetric images of the
 Circinus galaxy showing
a clear bipolar scattering cone in the J band. The south--east cone was
previously undetected at optical
wavelengths because it was hidden behind the heavy extinction of the galactic 
disk. At longer wavelengths, the H and K band images show more compact 
structures due to the dominance of dichroic absorption over scattered 
radiation.
We have successfully applied an adapted version of the Y95 
model to
interpret the observed polarized flux distribution. This model includes two
geometrical identical scattering cones, diametrically opposite to each other, 
with the forward cone in the south--east direction at a position angle of 
125$\deg$ and an opening half angle of 45$\deg$. The inclination of the system
to the line of sight is 50$\deg$. The estimated optical extinction 
A$_V$ to the 
nucleus through the torus is $>$66 mag. We estimate that the putative 
torus in the Circinus galaxy has an outer radius of $\sim$16pc.

\section{Acknowledgements}
M.R. thanks PPARC for support through a postdoctoral assistanship.
DMA thanks the TMR network (FMRX-CT96-0068) for a postdoctoral grant.
We thank Dolores P\'erez-Ram\'\i rez for her valuable help on the
production of this paper.

\vskip .3truecm
\leftskip=25pt
\parindent=-\leftskip

%\section{References:}

\bsp

\label{lastpage}

\end{document}